\documentclass{aa} 
\usepackage{txfonts}
\usepackage{graphicx}
\usepackage{natbib}

\bibpunct{(}{)}{;}{a}{}{,}

\newcommand{\FIG}[0]{}
\begin{document}
\title{Transverse stability of relativistic two-component jets}

  \titlerunning{Relativistic two-component jets}
  \authorrunning{Z. Meliani}
  \author{Z. Meliani
           \inst{1}, R. Keppens \inst{1,2,3}}

   \offprints{Z. Meliani}

\institute{FOM-Institute for Plasma Physics Rijnhuizen, Nieuwegein, The Netherlands
\and Centre for Plasma Astrophysics, K.U.Leuven, Belgium
\and Astronomical Institute, Utrecht University, The Netherlands \\
 \email{zakaria.meliani@obspm.fr, Rony.Keppens@wis.kuleuven.be}
 }

\date{Received ... / accepted ...}

\abstract{Astrophysical jets from various sources seem to be stratified, with 
a fast inner jet and a slower outer jet. As it is likely that the launching 
mechanism for each component is different, their interface will develop 
differential rotation, while the outer jet radius represents a second interface where disruptions may occur.}
{We explore the stability of stratified, rotating, relativistic two-component 
jets, in turn embedded in static interstellar medium.}
{In a grid-adaptive
relativistic hydrodynamic simulation with the AMRVAC code,
the non-linear azimuthal stability of two-component relativistic jets is 
investigated. We simulate until multiple inner jet rotations have been completed.}
{We find evidence for the development of an 
extended shear flow layer between the two 
jet components, resulting from the growth of a body mode in the inner jet, Kelvin-Helmholtz surface modes at their original interface, and their nonlinear interaction.
Both wave modes are excited by acoustic waves which 
are reflected between the symmetry axis and the interface of the two jet 
components. Their interaction induces the growth of near stationary, counterrotating vortices at the outer edge of the shear flow layer.
The presence of a heavy external jet allows to slow down their further development, and maintain a collimated flow.
At the outer jet boundary, small-scale Rayleigh-Taylor instabilities develop, without disrupting the jet configuration.}{We demonstrate that the cross-section of two-component relativistic jets, with a heavy, cold outer jet, is non-linearly stable.}

\maketitle

\section{Introduction}

Several astrophysical jet observations invoke a structured relativistic jet, 
in a direction perpendicular to the jet axis. This is the case for 
Active Galactic Nuclei (AGN) jets \citep{Siemiginowskaetal07}, with sometimes
clear evidence of a very fast, light jet and a heavy 
slow external outflow \citep{Girolettietal04}. 
Transversely structured, ultrarelativistic jet-like outflow is also proposed for Gamma Ray Bursts 
\citep{Rossietal02, Pengetal05}, to explain the break 
in the afterglow light curve. 

Theoretical studies for
a jet with two components can be found in various contexts, e.g. for the case 
of AGN in \cite{Sol89} and in \cite{Renaud98}, where an 
electron-positron central wind 
component is surrounded by an external, ideal MHD disk-driven jet. Another
two-component outflow model has been proposed by  
\cite{Bogovalov&Tsinganos05} where a relativistic pulsar wind is
surrounded and
collimated by an ideal MHD disk-driven wind.
Non-relativistic MHD scenarios with two-component jets are also suggested for classical
T-Tauri stars. In this case of Young Stellar Objects (YSO), two-component models were simulated in \cite{Melianietal06b}.
 There, the inner component extracts its energy from a hot corona
around the central region, while the second
component is launched from the thin accretion disk.  
Recent numerical simulations have progressed to general relativistic MHD (GRMHD) jet 
launching, as done by \cite{McKinney06} and \cite{Hardeeetal07}, and suggest a 
similar behavior with two component jets. \cite{Melianietal06a} have shown 
that a GRMHD jet launched from the inner region could be collimated by the pressure of external outflow from the disk.

The models of disk-driven jets show that while the mass loss from the accretion disk is higher than the
wind mass loss from the inner part, the jet speed increases near the axis, as all the models find that the asymptotic velocity of the jet
is close to the escape speed from the central engine. Indeed, there is a 
direct relation between the asymptotic speed and the depth of the gravitational
potential \citep{Mirabel99, Livio99}.  
Another piece of evidence from 
the models of jet launching that
the inner component of the jet is faster, is that the disk-driven jet mechanism is mainly an ideal MHD mechanism,
while the wind from the inner region could be more turbulent and is aided by 
non-ideal MHD processes
\citep{Melianietal06b}, hence
the inner jet can be accelerated to high Lorentz factor, even with a weak opening angle of 
the jet \citep{Spruitetal01}.
Therefore, a study of the transverse structure of two-component relativistic jets is a crucial step to 
understand the physics of high-speed jets and the mechanisms responsible in jet launching and 
collimation. In fact, the interactions between the two jet components may change substantially the dynamics and even question the stability of the relativistic jet.
Recent stability studies \citep{Hardee07} using the linearized RMHD
equations, can aid in identifying the normal modes for magnetized spine-sheath relativistic jets.

While most of these mentioned studies concentrate on magnetized jets, we here 
intentionally revisit non-linear stability issues for two-component, 
relativistic hydro jets.
The development of Kelvin-Helmholtz (KH) instabilities in 3D relativistic hydro jets has been 
investigated extensively \citep{Hardee00, Peruchoetal06}. 
Here, we study with unprecedented detail the effect of a transverse structure, consisting of an inner, fast, and faster rotating outflow, and an external,
`slow' (but relativistic), heavy, and slowly rotating jet, in turn embedded in a static medium.
This is done by means of a grid-adaptive computation focused on the transverse stability of a cross section of a jet with 
two components. We analyse the effect of density and rotation velocity changes 
at the interface between the two components in the jet morphology, and at the
outer boundary with the interstellar medium. 
As the mechanism responsible for the launching of each component 
is believed to be different, the energy flux and the angular momentum carried 
in each component is different. We will quantify the nonlinear
properties of the structured jet and show the ability of the heavy slow jet 
to collimate and stabilize the inner jet.
To capture all details, even with a grid-adaptive simulation, we perform
a 2$\frac{1}{2}$D hydrodynamic model, as a first step towards
more complex models in 3D, eventually to be augmented with magnetic fields.
Our relativistic hydro simulation uses a Synge-type equation of state,
taking full account of effectively varying polytropic index.

\section{Two-component jet model}

We set up a two-component structured jet 
with a 
typical total kinetic luminosity flux set to $L_{\rm jet, Kin} = 10^{46} {\rm ergs/s}$.
The external radius of the two-component jet is taken to be $R_{\rm out} \sim 0.1 {\rm pc}$, 
a value in accord with a jet opening angle $6^{\circ}$ at $1 {\rm pc}$ from the
jet source, which represent observed values for M87 \citep{Birettaetal02}.
We set the inner jet radius $R_{\rm in} =  R_{\rm out}/3$,
and we ensure that
this inner component carries only a small fraction $f_{\rm in}= 0.1\%$ 
of the total kinetic luminosity flux, 
while the external jet carries the remaining $f_{\rm out} = 99.9\%$.
For the initial condition, we adopt a uniform outflow velocity $v_{z}$ along the jet axis in each jet component, with 
$v_{\rm z, out} = 0.9428$ (hence $\gamma_{\rm z, out} \sim 3$, { since the speeds are normalised to the light speed}) 
 for the outer,
slow jet within $R_{\rm in} < R < R_{\rm out}$. This represents a reasonable value for relativistic jets at the pc scale.
The inner jet has a fast outflow set to $v_{\rm z, in} =0.9982$, 
with corresponding Lorentz factor $\gamma_{\rm z, in} \sim 16.6$.

Two-component jet models also indicate that the spin of the inner beam is 
higher than the spin of the external jet. 
Therefore, the initial rotation adopts two different profiles, 
one for the inner, and one for the external jet. This is inevitable when
these two jet components are launched from different regions and with 
different mechanisms.  
We choose for the external jet a radially decreasing rotation profile, with a 
decrease faster than Keplerian, as this component is launched from an
accretion disk and the external streamlines in the jet expand faster than the 
inner streamlines. Moreover, we assume that the angular momentum extracted 
along each streamline varies little, since we adopt a toroidal 
velocity varying as $1/R$. For the initial inner jet rotation profile, 
we take a toroidal 
velocity increasing with $\sqrt{R}$ from the axis. This toroidal velocity 
vanishes on axis and the equivalent classical centrifugal force given by this 
profile is constant. Summarizing, we take
\begin{equation}
v_{\varphi}=\left\{
\begin{array}{ccr}
V_{\varphi, \rm in} \sqrt{\frac{R}{R_{\rm in}}}& & R \le R_{\rm in}\,,\\
V_{\varphi, \rm out} {\frac{R_{\rm in}}{R}}& & R_{\rm in}<R<R_{\rm out} \,,
\end{array}
\right.
\end{equation}
such that we have a discontinuity in the toroidal velocity at the boundary between the two
jets when fixing $V_{\varphi,\rm in} = 0.05$ and $V_{\varphi, \rm out}=0.005$.
A faster rotating inner component could be due to the small expansion of 
the inner streamlines, since this component is confined by the external 
outflow. 
Also, a fast rotating inner jet can effectively extract the angular momentum 
from the 
central region and carry it away with a very small mass flux.
This rotation profile of the inner jet obeys the necessary (non-relativistic)
Rayleigh criterion 
for stability $v_{\varphi}\,r (r\,v_{\varphi})'>0$, which 
makes the inner jet centrifugally stable. The external outflow is marginally 
stable as $v_{\varphi}\,r (r\,v_{\varphi})'=0$.
However, the interface between the two components does not verify the Rayleigh 
criterion, so the shear flow interface is unstable.

The densities are also assumed to be constant throughout inner jet beam, outer 
jet beam, and external region. The external medium is hot and rarified 
and has a { small reference number density.}  The corresponding proper medium density $\rho_{\rm med}$ is in the computation used as a scaling value, together with $c=1$ and a unit length of $1$ pc. At the parsec scale,
the jet head has previously shocked the jet beam surroundings, so that the jet environment 
represents a relativistically hot, rarified, static external medium.
{ However, in this model we assume a number density for the external medium of $1 {\rm cm}^{-3}$,
which is somewhat higher than the values obtained in some simulations
for the cocoon surrounding the jet. However, the density of the external medium has little to no effect in the collimation of the 
jet cross section as the external jet has slow rotation and using smaller value for density in this simulation 
will only decrease the time step for integration.}
Values for the jet component proper densities are estimated from the following arguments prevailing at the jet head.
{ The propagation speed of the jet head is estimated
using momentum balance in a 1D approximation,} $ v_{\rm jet}^{\rm head}\,=\,v_{z, \rm out} \, {\sqrt{\eta_{R}}}/({\sqrt{\eta_{R}}+1})$ \citep{Martietal97}. In this 
expression,
$\eta_{R}=\gamma_{\rm out}^2 {\rho_{\rm out}h_{\rm out}}/{\rho_{\rm med}h_{\rm med}}$ is the ratio between the inertia of the (outer) jet and the external medium, and $h_{\rm out}$ is the enthalpy of the outer beam. Requiring that
this outer jet carries 
a kinetic luminosity flux of
$f_{\rm out} L_{\rm jet, Kin} =  \left(\gamma_{\rm out}\,h_{\rm out}-
1\right)\rho_{\rm out}\gamma_{\rm out}\pi
\,(R_{\rm out}^2-R_{\rm in}^2) v_{\rm out}$, we can deduce a relation for $h_{\rm out}$ 
and thus find the density $\rho_{\rm out}$, by prescribing a jet head 
Lorentz factor.
We assume that in the interaction with the external medium, the heavy, outer jet undergoes 
a deceleration to a Lorentz factor $\gamma_{\rm jet, out}^{\rm head} \sim 2.5$.
{ This value agrees with the fact that the jet, at the $\rm pc$ scale, 
is still relativistic, and that the outer jet interacts weakly with the external
medium.}
The resulting density is $\rho_{\rm out}\sim 270 \rho_{\rm med}$.
A similar argument for the inner, light, jet, which therefore must undergo a 
stronger deceleration, uses a Lorentz factor $\gamma_{\rm jet, in}^{\rm head} \sim 1.5$, and the density works out to be $\rho_{\rm in}\sim 0.2\rho_{\rm med}$. 

Finally, the pressure is deduced by supposing initial transverse equilibrium between 
the pressure gradient and the centrifugal force, by making the assumption of a constant (but different for inner versus outer jet) local `effective polytropic index'. Denoting the latter with $\Gamma_{\rm in}=4/3$ and $\Gamma_{\rm out}=5/3$ 
{ (the Synge equation of state is not used to determine this initial equilibrium)}, this works out to be
\begin{equation}
p=\left\{
\begin{array}{ccr}
-\frac{\Gamma_{\rm in}-1}{\Gamma_{\rm in}}\rho_{\rm in}+
\left(\frac{\Gamma_{\rm in}-1}{\Gamma_{\rm in}}\rho_{\rm in}+p_{0}\right)
\left(1-\frac{v_{\varphi}^2}{1-v_{\rm z,in}^2}\right)^{-\frac{\Gamma_{\rm in}}{\Gamma_{\rm in}-1}} & & \,,\\
{\rm for \,\,\,\,} R \le R_{\rm in}\,, & &\\
-\frac{\Gamma_{\rm out}-1}{\Gamma_{\rm out}}\rho_{\rm out}+
\left(\frac{\Gamma_{\rm out}-1}{\Gamma_{\rm out}}\rho_{\rm out}+p_{\rm in}\right)
\left(\frac{1-\frac{v_{\varphi}^2}{1-v_{\rm z, out}^2}}{1-\frac{V_{\varphi, \rm out}^2}{1-v_{\rm z, out}^2}}\right)^{\frac{\Gamma_{\rm out}}{2\,(\Gamma_{\rm out}-1)}}& &  \,, \\
{\rm for \,\,\, } R_{\rm in}<R<R_{\rm out} \,, & & \\
\end{array}
\right.
\end{equation}
where 
\begin{equation}
p_{\rm in}=-\frac{\Gamma_{\rm in}-1}{\Gamma_{\rm in}}\rho_{\rm in}+
\left(\frac{\Gamma_{\rm in}-1}{\Gamma_{\rm in}}\rho_{\rm in}+p_{0}\right)
\left(1-\frac{V_{\varphi, \rm in}^2}{1-v_{\rm z, in}^2}\right)^{-\frac{\Gamma_{\rm in}}{\Gamma_{\rm in}-1}} \,.
\end{equation}
We choose an initial pressure at the jet axis of $p_{0} = 10^{-4}{\rm \rho_{
\rm med} c^2}$. 

\begin{figure}
\begin{center}
\FIG{
{\rotatebox{0}{\resizebox{\columnwidth}{5cm}{\includegraphics{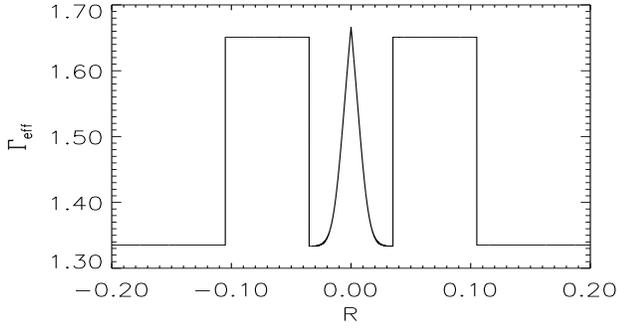}}}}
}
\caption{The effective polytropic index in the initial conditions in a 1D cut through the structured jet.}\label{LabelFig_JetCut1D}
\end{center}
\end{figure}

The governing equation of state is taken as a Synge-type relation, also used in
\cite{Melianietal04}.
As a result of the above profile prescriptions, the external jet is relativistically cold with an effective polytropic index $\Gamma_{\rm eff}=5/3$ and the 
inner jet has its effective
polytropic index stratified, since the pressure increases fast from the axis where 
$\Gamma_{\rm eff}=5/3$, to the inner jet boundary where $\Gamma_{\rm eff}=4/3$.
In Fig.~\ref{LabelFig_JetCut1D}, we show the effective polytropic index of the
initial condition.
With the ratio between the inertia of the outer jet component and the outer shell
of the inner jet, i.e. $\gamma_{\rm out}^2\rho_{\rm out}h_{\rm out}/{\gamma_{\rm in}^{2}\rho_{\rm in}h_{\rm in}}$ of about $1/2$,  the local sound speed $c_s$
is very high in the inner jet and this makes this shell subject to a strong 
interaction with the outer jet.
Moreover, with this model, the angular momentum carried by the 
inner jet is of the same order as 
the angular momentum carried by the external jet.
The outer shell of the inner jet is subsonically rotating with a speed of 
order $\sim 0.1 c_s$, while the outer, faster rotating jet is actually rotating at $\sim 0.03 c_s$, due to the different thermodynamic
conditions. Such configuration is favourable to asymmetric instabilities and hence their growth times will be of order of the radial sound-crossing time \citep{Hardee04}.

The computational domain of this simulation is a 2D box of size $-0.2 {\rm pc} <x<0.2 {\rm pc}$ and $-0.2 {\rm pc} <y<0.2 {\rm pc}$, 
and
the simulation is performed in Cartesian coordinates with the HLLC flux 
formula \citep{Mignone&Bodo05}. 
The simulation is ran till time $t=22.75$. Since the time for a total
$2\pi$ rotation of the external boundary of the inner jet is $4.4$,
the simulation follows about $5$ full rotations of the inner jet.{ The
corresponding distance of jet propagation of the jet beam during this time is 
$21.45 {\rm pc}$}.
The simulation is done using the AMRVAC code with Synge equation of state,
using the hybrid block grid adaptivity approach from \cite{Bartkep07}.
We take a base resolution of $128\times 128$, allowing for 5 grid levels,
reaching effective resolution of $2048^2$.

\section{Nonlinear evolution and stability}

We first did a 1D run, only evolving a radial cut along the cylindrical radius 
through the jet.
This 1D simulation demonstrated that the two-component jet remains collimated and keeps its internal two-component structure. Because of the initial 
discontinuity between the two jet components, acoustic waves are created there which propagate through the
two jet components. The acoustic waves propagating in the inner jet meet at the axis and are reflected. The waves propagating in the external jet get partially
reflected on the external boundary of the jet. 
However, these acoustic waves do not 
disturb the two-component jet structure. After three full
rotations, the inner jet settles into a stationary state with somewhat higher density on axis, as mediated by the propagating acoustic waves. As the outer jet component has a high inertia and low sound speed,
it efficiently reflects acoustic waves at the inner
boundary between the two components. 
 
In 2D, the simulation shows in addition, the development of surface and body normal modes. Prior to $t\sim 1$, the shear flow interface develops azimuthal Kelvin-Helmholtz 
instabilities due to the velocity gradient. Meanwhile, initially axisymmetric 
acoustic waves reflect on the jet axis (as in the 1D run) and then give rise to the growth of a body mode in the inner jet.
The latter shows a spiral structure with a dominant azimuthal mode number $m=4$ character. 
Both modes grow and extract energy 
mainly from the inner jet.
As mentioned before, the sound speed in the external jet is rather 
slow $c_{\rm s, out}\sim 0.15 c$, and four times less than the sound speed in 
the inner jet. The body mode in the inner jet thus grows fast, 
as its growth is related to the travel time taken by acoustic waves across the 
inner jet. In such a situation, the body mode dominantly disturbs 
the inner jet component, and the inner boundary of the jet where the azimuthal surface mode exists. This strongly couples the two modes, and from $t\sim 1.2$ onwards the body mode will 
contribute to surface mode development, and induce spiral deformations which penetrate as 4 `arms' into the outer jet component. This causes the jet to expand. In this expansion phase,  the outer surface of the two-component jet becomes unstable to a Rayleigh-Taylor-type 
instability. This starts at about $t\sim 3.7$ and leads to small-scale structure. However, for the entire simulated time-interval,  this does not disrupt the overall jet structure dramatically. 
\begin{figure}
\begin{center}
\FIG{
{\rotatebox{0}{\resizebox{\columnwidth}{5cm}{\includegraphics{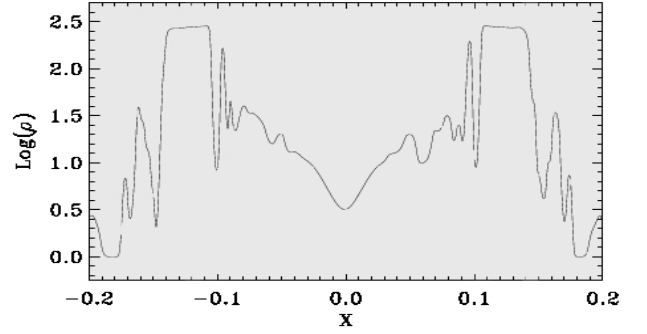}}}}
}
\caption{Density in a 1D cut through the structured jet after 5 full rotations of the inner jet .}\label{LabelFig_JetCut1Dt5}
\end{center}
\end{figure}

The non-linear interaction of the two modes in the inner part eventually causes 
the development of an extended shear flow shell, separating the inner 
and the outer jet.
This shear flow broadens and compresses 
the inner jet and thereby increases its density, while decelerating it in the 
vertical (axial) direction due to mass conservation.  
During the formation of the shear flow layer, at about $t\sim 7$ the body mode changes
from a dominant azimuthal mode number $m=4$ character, to an 
elliptic $m=2$. Eventually, as in the 1D run,  when the acoustic waves in the inner jet all but disappear, the inner jet becomes axisymmetric again and a more
stationary state is reached (at about $t\sim 8$), where the radius of the inner jet drops to 
$R_{\rm in} \sim 0.015 {\rm pc}$ while the density at the axis reaches 
$\rho_{\rm in}\sim 3.17 \rho_{\rm med}$. From about this time,
acoustic waves propagating in the extended shear
flow region get reflected at its bounding interfaces,
as the layer becomes a kind of resonant cavity. 
Throughout, a fraction of the angular momentum gets transferred to the inner jet region
as we find an acceleration of the flow rotation to $v_{\rm \varphi}\sim 0.076 c$ 
at radius $R_{\rm in}$, which is needed to radially support the increased
pressure of the more turbulent shear flow region. This transfer is mostly { associated with the `arms' extending from the 
rotation axis to the outer part of the inner jet, and transfer happens in accord with the sound speed}. 
{ Moreover,
 the compression of the inner jet contributes to the acceleration of the flow rotation.}
However, the simulation shows that there is no transfer of angular momentum 
from the inner jet to the outer jet component, since its rotation speed does 
not increase. 
All the angular momentum and energy extracted from the inner jet ends up in the shear region.
 
\begin{figure}
\begin{center}
\FIG{
{{\resizebox{7.0cm}{5.8cm}{\includegraphics{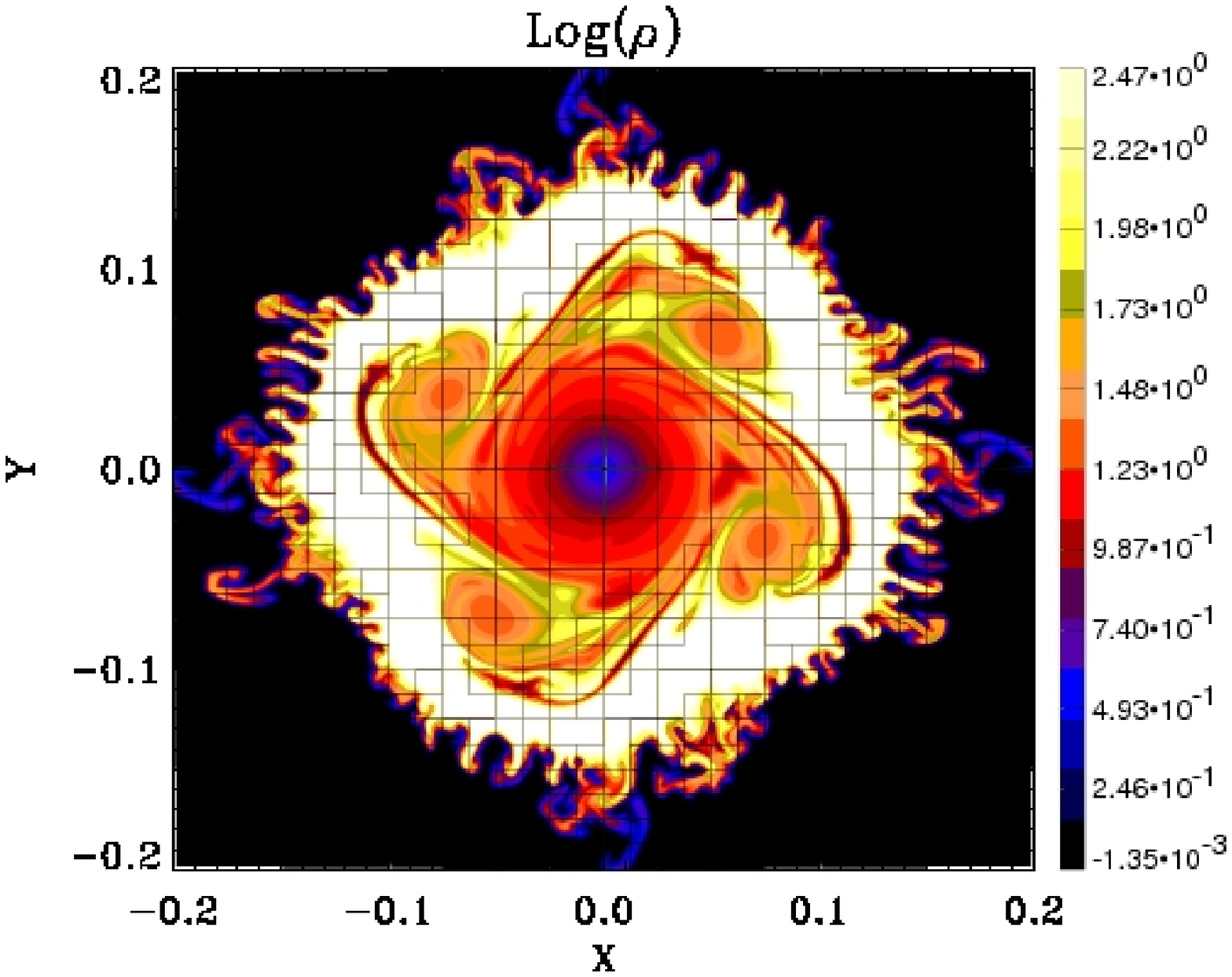}}}
 {\resizebox{7.0cm}{5.8cm}{\includegraphics{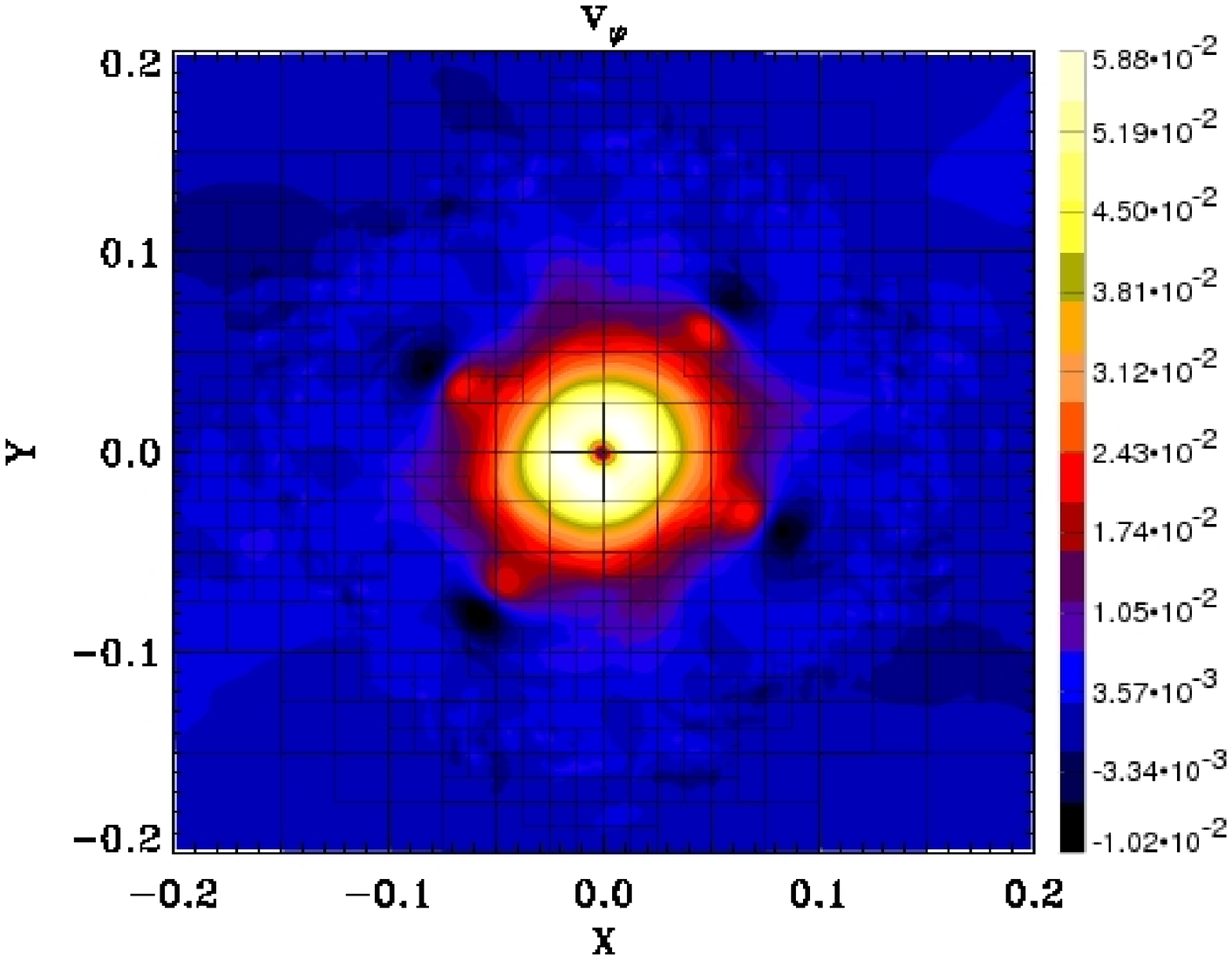}}}}
{{\resizebox{7.0cm}{5.8cm}{\includegraphics{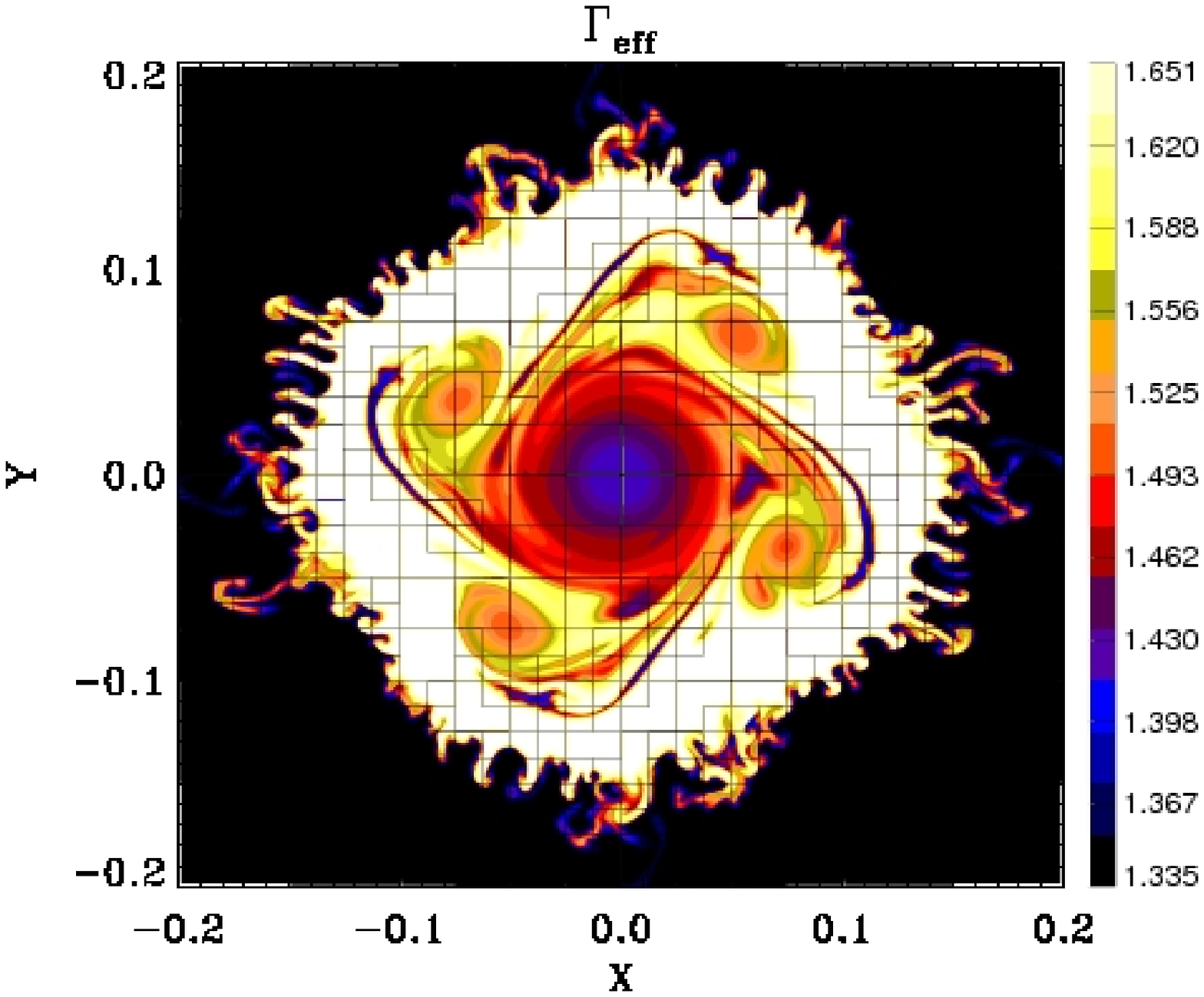}}}
 {\resizebox{7.0cm}{5.8cm}{\includegraphics{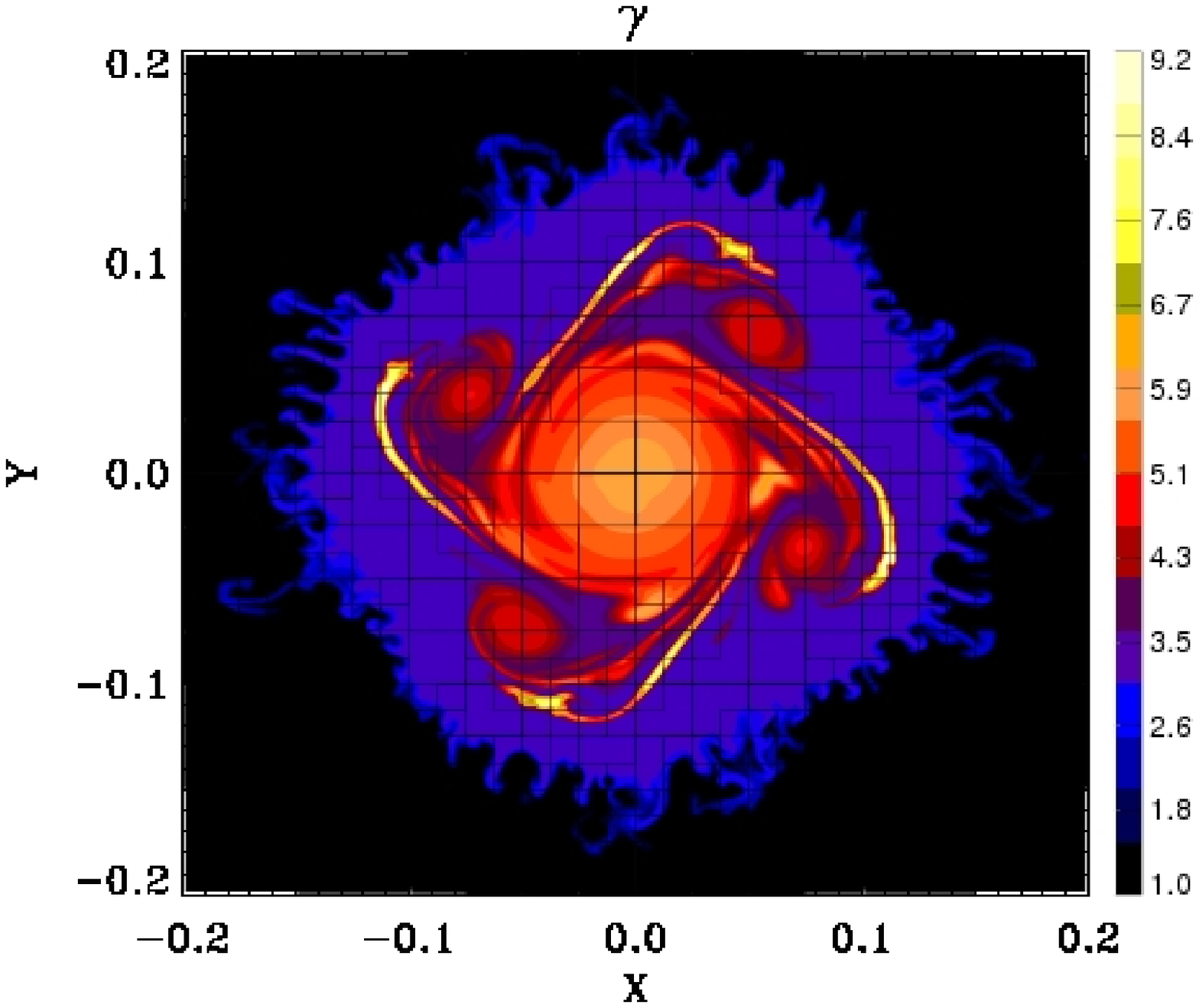}}}}
}
\caption{Density, toroidal velocity, effective polytropic index, and 
Lorentz factor for the two-component jet, after 5 full rotations of the inner jet.}\label{LabelFig_JetCut2D}
\end{center}
\end{figure}

Fig.~\ref{LabelFig_JetCut2D} shows various quantities at the end time $t=22.75$,
and it can be seen how
$4$ `arms' have penetrated into the outer jet region.
The shear flow region then extends between $ 0.015 {\rm pc}$ and $0.1 {\rm pc}$, as seen in $v_{\varphi}$ Fig.~\ref{LabelFig_JetCut2D}, 
{ and also in density $\rho$ in the 1D cut  through the structured jet Fig.~\ref{LabelFig_JetCut1Dt5}}.
The pressure and the centrifugal force in the sheared shell increases, and together with the ram pressure associated with the protruding `arms', the outer jet
region has expanded radially. 
The radius of the outer jet therefore rises to 
$R_{\rm out}\sim 0.15 {\rm pc}$.
However, the extension of the `arms' into the outer jet region stops when 
they reach a force equilibrium with the external jet, which compresses them. 
In the process, a series of stationary counterrotating vortices develop at the outer edge of the shear flow region. 
These vorticity wave patterns extract energy from the inner jet
and from waves reflected at both bounding interfaces of
the formed shear flow region.
In the simulation, we find that the spiral arms and the vorticity 
waves are coupled, and the vortices remain near stationary.

The thermal energy in the inner jet remains relativistic, with an effective 
polytropic index $\Gamma_{\rm eff}\sim 1.38$, while in the outer jet thermodynamic conditions remain classical with an effective polytropic 
index $\Gamma_{\rm eff}\sim 1.652$.
However, the thermal energy in the shear flow layer, and in the vortex structures is mildly relativistic. In the shear flow the effective polytropic 
index is of the order $\Gamma_{\rm eff}\sim 1.45$, and in the center of the
vortex we find $\Gamma_{\rm eff}\sim 1.5$.

\section{Conclusions}

We investigated the dynamical consequences of an inner fast, light, and fast 
rotating outflow, on a slower, heavy and slower rotating relativistic jet. The
latter could be launched from an accretion disk surrounding the engine for the inner jet.
We performed a numerical simulation to study the azimuthal stability of the 
cross section of this two component jet.
The massive, cold outer jet makes the inner, light jet act as a resonant cavity 
due to strong reflection of acoustic waves at the interface. This induces various linear instabilities which interact strongly.
The simulation shows evidence for a transition to a near-stationary state, three-component
state, with a separating shear flow region between the two components. 
In the process, stationary counterrotating vortices develop. 
Their growth
gives rise to significant thermal and ram pressure in the shear region,
which compresses the inner jet and partially expands the external jet. 
However, the heavy, cold outer jet slows down the instability development,
and the overall increased cross section remains collimated after 
several dynamical rotation times of the inner component. 
The resulting more turbulent shear layer between the two jet components 
could be responsible for continuous in-situ reacceleration of electrons, 
and have a measurable impact on the radiation spectra, as already suggested 
by \cite{Stawarz&Ostrowski02}.

{ The non linear stability analysis of the jet cross section presented here allows 
to investigate precisely the development of azimuthal instabilities and 
their influence on the inner and the external structure of the jet cross 
section. The azimuthal instabilities between the two jet components ultimately 
modify the speed of the inner jet and the width of the external jet which should influence
the 3D structure of the jet.}
In future work, we will study the interaction between two components in full 
3D simulations. They should explore the influence of the initial rotation 
profile { and the relative influence of azimuthal versus longitudinal 
instabilities for realistic multi-component jets}.
 Finally, magnetic fields must be included for quantifying their further
role in relativistic structured jet collimation.

\begin{acknowledgements}
We acknowledge financial support from the Netherlands Organization for Scientific Research, NWO-E grant 614.000.421, and computing resources supported by
NCF. Part of the computations made use of the VIC cluster at K.U.Leuven.
\end{acknowledgements}

\bibliographystyle{aa}

\end{document}